# Dimensional crossover and symmetry transformation of the charge density waves in VSe$_2$

P. Chen,[1#*] Y.-H. Chan,[2,3#] R.-Y. Liu,[4,5,6] H. T. Zhang,[1] Q. Gao,[1] A.-V. Fedorov,[7] M. Y. Chou,[2,8,9*] and T.-C. Chiang[4,5*]

[1]Key Laboratory of Artificial Structures and Quantum Control (Ministry of Education), Shenyang National Laboratory for Materials Science, Shanghai Center for Complex Physics, School of Physics and Astronomy, Shanghai Jiao Tong University, Shanghai 200240, China

[2]Institute of Atomic and Molecular Sciences, Academia Sinica, Taipei 10617, Taiwan

[3]Physics Division, National Center for Theoretical Sciences, Taipei 10617, Taiwan

[4]Department of Physics, University of Illinois at Urbana-Champaign, 1110 West Green Street, Urbana, Illinois 61801-3080, USA

[5]Frederick Seitz Materials Research Laboratory, University of Illinois at Urbana-Champaign, 104 South Goodwin Avenue, Urbana, Illinois 61801-2902, USA

[6]National Synchrotron Radiation Research Center, 101 Hsin-Ann Road, Hsinchu Science Park, Hsinchu 30076, Taiwan

[7]Advanced Light Source, Lawrence Berkeley National Laboratory, Berkeley, California 94720, USA

[8]Department of Physics, National Taiwan University, Taipei 10617, Taiwan

[9]School of Physics, Georgia Institute of Technology, Atlanta, GA 30332, USA

Abstract

Collective phenomena in solids can be sensitive to the dimensionality of the system; a case of special interest is VSe$_2$, which shows a $(\sqrt{7} \times \sqrt{3})$ charge density wave (CDW) in the single layer with the three-fold symmetry in the normal phase spontaneously broken, in contrast to the $(4 \times 4)$ in-plane CDW in the bulk. Angle-resolved photoemission spectroscopy (ARPES) from VSe$_2$ ranging from a single layer to the bulk reveals the evolution of the electronic structure including the Fermi surface contours and the CDW gap. At a thickness of two layers, the ARPES maps are already nearly bulklike, but the transition temperature $T_C$ for the $(4 \times 4)$ CDW is much higher than the bulk value of 110 K. These results can be understood as a result of dimensional crossover of phonon instability driven by a competition of nesting vectors. Our study provides key insights into the CDW mechanisms and offers a perspective in the search and control of emergent phases in quantum materials.

Collective behavior of the coupled electronic and lattice degrees of freedom in connection with CDW formation is an actively explored area. It is of broad interest as CDWs can entangle or compete with other ordering effects such as superconductivity and magnetism [1-3]. Transition metal dichalcogenides (TMDCs) made of a stack of van-der-Waals-bonded molecular layers provide an excellent platform for exploring dimensional effects on spontaneous charge ordering and other related properties including the quantum spin Hall effect, metal-insulator transition, etc., as these materials can be readily prepared as ultrathin films with a thickness down to the single layer [4-9]. Such dimensional effects are of basic interest and can be utilized for property tuning or control via a thickness-dependent Lifshitz transition [10].

While CDWs have been investigated for decades, there is not yet a definitive universal model explaining all observed cases due to possible competition and entanglement of electron-electron and electron-phonon interactions [11-14]. $VSe_2$ is of special interest due to its unusual $(\sqrt{7} \times \sqrt{3})$ CDW symmetry in the single layer [15-17], which is distinct from the in-plane $(4 \times 4)$ CDW symmetry in the bulk [18-22]. This thickness-mediated CDW symmetry transformation is puzzling and unexpected because of the quasi-two-dimensional nature of the $VSe_2$ crystal structure. $VSe_2$ thus offers a unique test case for the various models or theories for CDW formation. For bulk $VSe_2$, prior experiments suggested three-dimensional (3D) Fermi surface nesting as a possible mechanism for the CDW transition [20, 21], and recent theoretical calculations revealed imaginary phonon frequencies relevant to the transition [23]. Our layer-resolved ARPES measurements, reported herein, directly probe the effects of dimensionality on the electronic structure in the ultrathin film limit. Subtle changes in the Fermi surface contours as a function of film thickness are observed; the resulting gentle shift of the nesting condition nevertheless drives different phonon instabilities. In conjunction with theory, our observations uncover the underlying reasons for the dimensional crossover of CDW properties including the transformation of the CDW lattice symmetry.

ARPES spectra for $N$ = 1-3 trilayers (TL) and bulk ($N = \infty$) VSe$_2$ along the $\overline{\Gamma M}$ direction are presented in Figs. 1(a) and 1(b) for the normal phase at 300 K and the CDW phase at 10 K, respectively. The sharpness of the bands indicates excellent sample quality. The experimental dispersion relations and the multiplication of bands (formation of quantum well states, most easily seen around the zone center) as $N$ increases from 1 to 3 are in excellent agreement with theoretical results for the normal phase [Fig. 1(c)]. This establishes the layer-by-layer buildup of the film in the experiment. The theoretical spectral weights of the bands [Fig. 1(c)], indicated by the blue and red circles for the V 3$d$ and Se 4$p$ states, respectively, reveal that the bands near the Fermi level, relevant to the CDW transition, are dominated by the V 3$d$ states. All samples, $N$ = 1-3 and $\infty$, are metallic in the normal phase, with bands crossing the Fermi level. Comparing the results in Figs. 1(a) and 1(b), changes in the ARPES data are subtle when the samples transition from the normal phase at 300 K into the CDW phase at low temperatures. The main differences are sharper spectral features at 10 K due to reduced thermal scattering. No band folding effects are apparent, indicating a weak superlattice distortion in the CDW phase. An extra very faint band observed along $\overline{\Gamma M}$ between -1.5 and -1 eV for the 1-TL sample, not seen in the theoretical band structure, is derived from the Se 4$p$ bands associated with 30°-rotated minority domains in the film. This feature also exists at 300 K in the normal phase and has no relation to the CDW transition.

Measured Fermi contour maps for $N$ = 1-3 and $\infty$ are shown in Figs. 2(a) and 2(b) for sample temperature at 300 and 10 K, respectively. For comparison, calculated Fermi contour maps for the normal phase are shown in Fig. 2(c), which are fairly close to the experimental results for the normal phase at 300 K. For $N = 1$, the experiment contours consist of six electron pockets radially distributed around the zone center. Each pocket, centered about a $\overline{M}$ point, assumes a simple racetrack shape (a distorted ellipse with long parallel sides, or "straightaways"). The two straightaways of each pocket are conducive to nesting, thus facilitating CDW formation. Indeed, dark spots develop around the pocket edges for the CDW phase at 10 K for $N = 1$, which are indicative of gap formation at the Fermi level. The resulting CDW phase, with a $(\sqrt{7} \times \sqrt{3})$ structure [15], breaks the three-fold symmetry of the system. The $(1 \times 1)$ Brillouin zone (black hexagon), the CDW lattice vectors $\mathbf{q}_1$ and $\mathbf{q}_2$, and some repeated CDW zones (red dotted lines) are indicated in Fig. 2(c). Vector $\mathbf{q}_1$, represented by a red double-ended arrow in Fig. 2(a), closely spans the two straightaways of each racetrack and is relevant to the nesting.

For $N = 2$, the experimental Fermi contour map at 300 K is overall similar to the $N = 1$ case, but the width of the electron pocket becomes slightly smaller. At 10 K, no dark spots are observed, but the intensity of the two straightaways of each pocket becomes weaker. This suggests a diffuse gap opening over a background. Similar behavior is also observed for $N = 3$, although the racetrack pocket become even narrower. This trend continues to the bulk case, which is known to adopt a $(4 \times 4 \times 3)$ symmetry [18]. The changeover from a $(\sqrt{7} \times \sqrt{3})$ CDW at $N = 1$ to a $(4 \times 4)$ in-plane CDW happens already at $N = 2$ based on STM measurements at low temperatures [24]. The $(4 \times 4)$ CDW lattice vectors $\mathbf{p}_1$ and $\mathbf{p}_2$ and the corresponding CDW repeated zones (red dotted lines) are indicated in Fig. 2(c) for $N = 2$, 3, and $\infty$. The blue double-headed arrows in Fig. 2(a), representing $\mathbf{p}_{1,2}$, also nearly span the two straightaways of each racetrack pocket. Evidently, the $(\sqrt{7} \times \sqrt{3})$ and $(4 \times 4)$ nesting conditions are both nearly satisfied.

ARPES maps taken at 300 K along the $\overline{K}$-$\overline{M}$-$\overline{K}$ direction for $N$ = 1-3 and $\infty$ are shown in the upper row in Fig. 3(a); a V-shaped band centered about $\overline{M}$, mainly derived from the V $3d$ states, is seen. Corresponding ARPES maps symmetrized about the Fermi level are presented in the lower row in Fig. 3(a). Similar results obtained at 10 K [Fig. 3(b)], in comparison with the 300 K data, demonstrate CDW gap formation at the Fermi level for $N$ = 1-3 but not for $N$ = $\infty$. The data are analyzed by curve fitting of the energy distribution curves with a phenomenological self-energy expression [25], and the gaps extracted as a function of temperature are shown in Fig. 3(c). The absence of a clear ARPES gap for the bulk case, consistent with prior STM and ARPES studies [16, 22, 26], can be attributed to the 3D nature of the CDW in the bulk. ARPES, with a poor $k_z$ resolution relative to the short Brillouin zone dimension along $z$, does not reveal the 3D mini gaps. Excluding the bulk case, the measured gap squared as a function of $T$ is fitted with a BCS mean field gap equation for each data set [Fig. 3(c)] [8]. The extracted zero-temperature gap $\Delta(0)$ is shown in Fig. 3(d) for the different cases. The CDW transition temperatures $T_C$ extracted from the same fitting (220, 180, and 170 K for $N$ = 1-3, respectively) are plotted in Fig. 3(e), and the known bulk $T_C$ = 110 K [18] is included for completeness. $T_C$ in the 3 TL film still does not quite converge to the bulk value, which implies a substantial 3D character of the bulk CDW. It is interesting to note that the same trend of a higher $T_C$ in the 2D limit has been observed for TiSe$_2$, TaSe$_2$, and NbSe$_2$ [6-8, 27]. However, VSe$_2$ is unique for which the CDW symmetry is distinct at $N$ = 1.

Fermi surface nesting in 2D involves reciprocal vectors spanning parallel segments of Fermi contours, which can lead to gap opening and energy lowering. Figure 3(f) shows the extracted dispersion relations at 300 K for the V-shaped ARPES bands in Fig. 3(a) in comparison with $\mathbf{q}_1$ (red arrow), which is related to the ($\sqrt{7} \times \sqrt{3}$) CDW, and the projection of $\mathbf{p}_1$ (blue arrow) along the $\overline{\text{K}}$-$\overline{\text{M}}$-$\overline{\text{K}}$ direction (perpendicular to the straightaways), which is related to the (4 × 4) CDW. The top opening of each V shape at the Fermi level corresponds to the perpendicular distance between the two straightaways of the Fermi pocket. It matches fairly well with $\mathbf{q}_1$ at $N = 1$ and almost exactly with the projection of $\mathbf{p}_1$ for the bulk case, respectively. This changeover of nesting condition can drive the transition from the ($\sqrt{7} \times \sqrt{3}$) CDW at $N = 1$ to the (4 × 4) CDW at larger film thicknesses. The percentage mismatch for the two nesting conditions is shown in Fig. 3(g) for the different film thicknesses to illustrate the trend. For $N = 2$ and 3, both nesting conditions compete; experimentally, the (4 × 4) structure is the winner.

Computed phonon dispersion relations for the normal phase are shown in Fig. 4(a). For $N = 1$, imaginary frequencies are observed at $\mathbf{q}_1$ and $\mathbf{p}_{1,2}$, which correspond to electronically-driven instabilities toward ($\sqrt{7} \times \sqrt{3}$) and (4 × 4) distortions, respectively, but the instability at $\mathbf{q}_1$ for the ($\sqrt{7} \times \sqrt{3}$) distortion is more prominent and wins the competition. For $N = 2$ and 3, the instability at $\mathbf{p}_{1,2}$ becomes more prominent, which would favor the (4 × 4) distortion, in agreement with the observed symmetry switching between $N = 1$ and 2. For the bulk case, imaginary phonon modes are not observed along Γ-M-K-Γ with $k_z = 0$ as plotted in Fig. 4(a), but can be found elsewhere in the bulk Brillouin zone. The distribution of the most prominent imaginary phonon frequencies [Fig. 4(c)] shows six symmetry-related regions of lattice instability. Three of these are centered at (1/4, 0, 2/3) and its three-fold partners in normalized hexagonal coordinate units within the Brillouin zone. The other three are centered at (0, 1/4, 1/3) and its three-fold partners. These results are further illustrated by the phonon dispersions plotted in Fig. 4(b) along the Γ-M-K-Γ in-plane directions but with $k_z$ set to 2/3 in normalized coordinate units. Clearly, the lattice instability is dominated by a mode at $\mathbf{p}_{1,2} = (1/4, 0)$ and $k_z = 2/3$. A vertical line trace along (1/4, 0, $k_z$) confirms that the instability is centered at $k_z = 2/3$ [Fig. 4(d)].

Altogether, the coordinates of the six lattice instability regions in the Brillouin zone agree with a $(4 \times 4 \times 3)$ symmetry for the CDW lattice distortion.

Our investigation shows that the fundamental driver for the CDW transitions in $VSe_2$ films is Fermi surface nesting, which by linear response leads to first-order energy lowering upon lattice distortion in accordance with the nesting condition. $VSe_2$ is nominally quasi-2D, and the measured Fermi surface map evolves only mildly as the film thickness increases. The main feature of the Fermi surface is a set of six racetracks, for which the opposing straightaways of each racetrack provide one of the two conditions for 2D nesting. The other condition is to match the straightaway separation with a superlattice vector, which is approximately satisfied by the $(\sqrt{7} \times \sqrt{3})$ and $(4 \times 4)$ spanning vectors, with the former preferred for $N = 1$, and the latter nearly perfect for $N = \infty$ as a result of the reduction of the Fermi pocket dimension as $N$ increases. Experimentally, the transition from the $(\sqrt{7} \times \sqrt{3})$ to $(4 \times 4)$ symmetry happens between $N = 1$ and 2; this transformation is rather unique among the large number of TMDCs exhibiting CDW transitions. A quantitative confirmation of the changeover in symmetry preference is provided by first-principles calculations of the phonon instabilities. While $VSe_2$ is nominally quasi-2D, its $(4 \times 4 \times 3)$ CDW in the bulk with a triple layer modulation implies a significant range of interlayer interaction along the layer normal. Indeed, the computed bulk band structure [28] shows substantial energy dispersions along the layer normal direction for some of the bands; this interlayer interaction leads to the symmetry changeover. Electron-correlation effects have often been invoked for explaining CDW formation. Such effects are ubiquitous in solids, and are in fact included in our first-principles calculations within the density-functional formalism. For the present case, there are no obvious electronic and structural features directly connecting electron correlation effects with the symmetry changeover between $N = 1$ and 2. Our results thus provide a clear-cut case for Fermi-surface nesting as the basic driver for the CDW order, but electron correlation effects could affect or modulate the detailed energetics. The measured CDW $T_C$ as a function of $N$ shows an interesting decreasing trend for increasing $N$. This can be understood on general grounds in terms of increased layer-to-layer fluctuations in thicker films, which tend to suppress long-range order. Putting all together, our detailed experimental and theoretical analysis of VSe$_2$ establishes a case that can be well understood in terms of the prevailing theory of electronic structure. This system has sufficient complexity (dimensional crossover of CDW symmetry) to offer a stringent test that challenges the premises of the various CDW theories.

**Acknowledgments**

We thank Prof. Q. Liang for helpful discussions. This work is supported by the Ministry of Science and Technology of China under Grant No. 2021YFE010160, the Science and Technology Commission of Shanghai Municipality under Grant No. 21JC1403000 (PC), the U.S. Department of Energy, Office of Science, Office of Basic Energy Sciences, Division of Materials Science and Engineering, under Grant No. DE-FG02-07ER46383 (TCC), and Academia Sinica (YHC and MYC). The Advanced Light Source is supported by the Director, Office of Science, Office of Basic Energy Sciences, U.S. Department of Energy, under Contract No. DE-AC02-05CH11231. Y.-H.C. acknowledges support by the Ministry of Science and Technology, National Center for Theoretical Sciences (Grant No. 110-2124-M-002-012) and National Center for High-performance Computing in Taiwan.

#These authors contributed equally：P. Chen, Y.-H. Chan.

*email: pchen229@sjtu.edu.cn;

mychou6@gate.sinica.edu.tw;

tcchiang@illinois.edu

FIG 1. ARPES data and band structures for $N$-layer VSe$_2$, with $N = 1$-$3$ and $\infty$. ARPES maps along $\overline{\Gamma M}$ taken at (a), 300 K and (b), 10 K. (c), Corresponding calculated band dispersions for the normal phase. The band characters, V $3d$ or Se $4p$, are color coded.

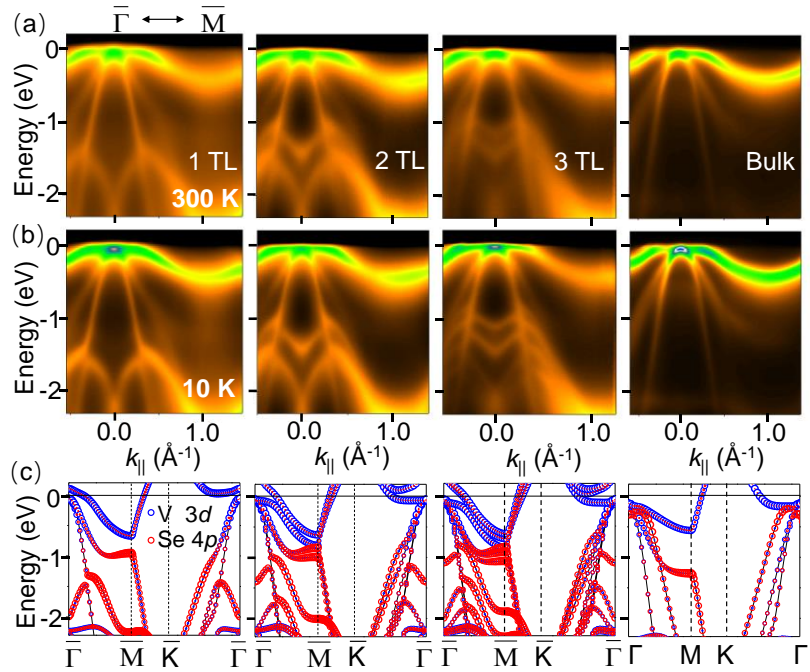

FIG 2. Fermi surfaces of *N*-layer VSe$_2$ in the CDW and normal phases. (a), Measured Fermi surface maps in the normal phase at 300 K obtained by integrating the ARPES intensity over ±10 meV about the Fermi level. The light blue hexagons indicate the first Brillouin zone. The red and blue arrows correspond to $\mathbf{q}_1$ and $\mathbf{p}_{1,2}$ for the $(\sqrt{7} \times \sqrt{3})$ and $(4 \times 4)$ CDW wave vectors, respectively. (b), Measured Fermi surface maps for the CDW phase at 10 K. (c), Calculated Fermi surface contours in the normal phase. Brillouin zones for the normal phase are outlined as black hexagons. Vectors $\mathbf{q}_{1,2}$ and $\mathbf{p}_{1,2}$ corresponding to the $(\sqrt{7} \times \sqrt{3})$ and $(4 \times 4)$ CDW primitive lattice vectors are indicated. The corresponding CDW repeated zones are indicated by red dotted lines.

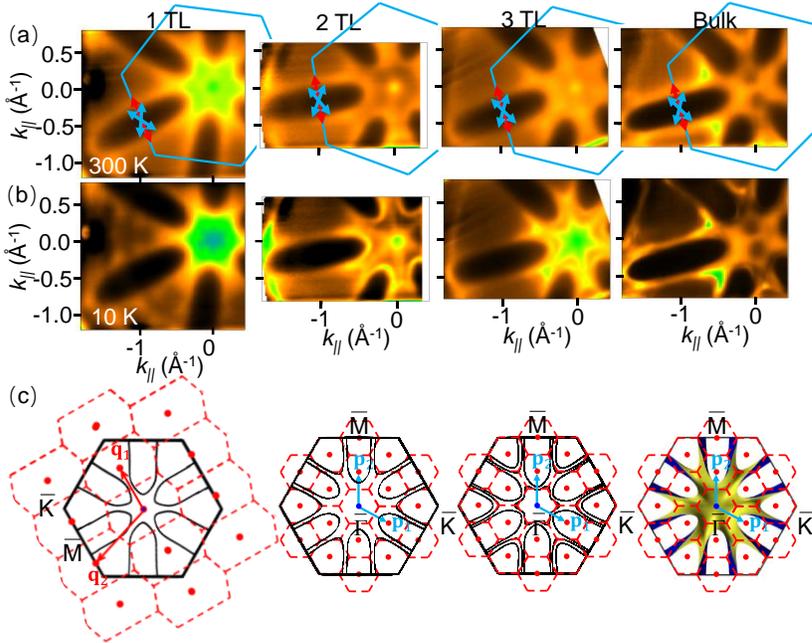

FIG 3. CDW gaps, transition temperatures, and nesting conditions. (a), ARPES spectra along the $\overline{K}$-$\overline{M}$-$\overline{K}$ direction for $N = 1\text{-}3$ and $\infty$ at 300 K (upper row) and the same spectra symmetrized in energy about the Fermi level (lower row). The symmetrization is to simplify the process of determining the gaps by canceling out the effects of Fermi-Dirac distribution (thermal broadening of the Fermi edge at finite temperatures). (b), Same as (a) but for data taken at 10 K. Gaps at the Fermi level are evident for $N = 1\text{-}3$. (c), Temperature dependence of the measured ARPES gaps, squared. The curves are fits, and the extracted transition temperature $T$c is labeled for each case. (d), Extracted gaps at $T = 0$. (e), Extracted transition temperatures $T$c. (f), Band dispersions along $\overline{K}$-$\overline{M}$-$\overline{K}$ extracted from fitting to the data for the normal phase at 300K. The nesting vectors $\mathbf{q}_1$ and $\mathbf{p}_1$ projected along the $\overline{K}$-$\overline{M}$-$\overline{K}$ direction are shown as a red arrow and a blue arrow, respectively. (g), Nesting vector mismatch as a function of film thickness evaluated for $\mathbf{q}_1$ and projected $\mathbf{p}_1$.

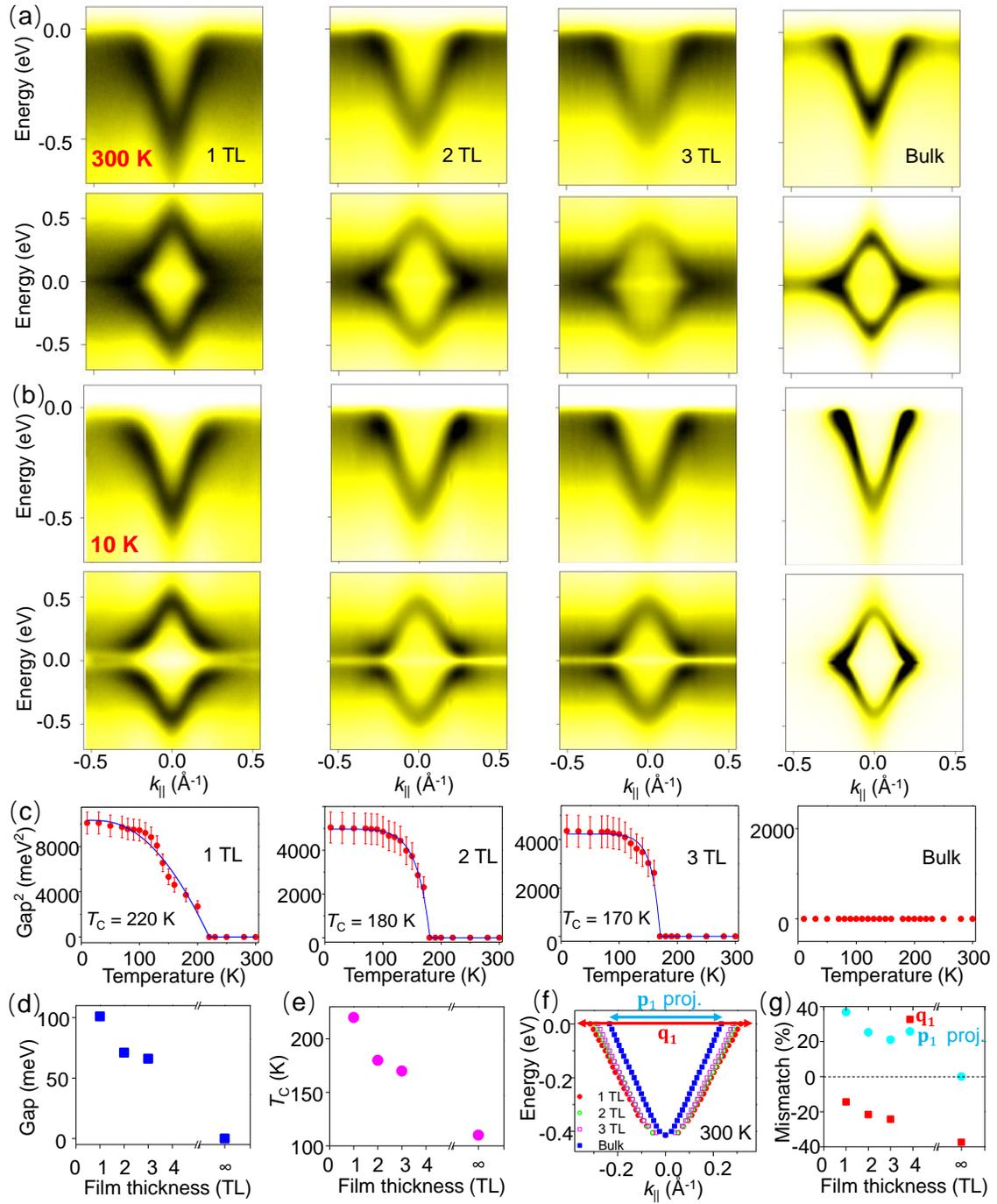

FIG 4. Phonon dispersions. (a), Calculated phonon dispersions for the normal phase for $N$ = 1-3 and bulk. Competing CDW wave vectors $\mathbf{q}_{1,2}$ and $\mathbf{p}_{1,2}$ are indicated. (b), Calculated phonon dispersions for bulk VSe$_2$ along the same in-plane directions Γ-M-K-Γ as in (a) but with $k_z$ set to 2/3 in normalized coordinate units to highlight the major imaginary phonon at (1/4, 0, 2/3). (c), Distribution of the most prominent imaginary phonon frequencies in $k$ space. (d), Phonon dispersion relations along a vertical line (1/4, 0, $k_z$). It shows the same major imaginary phonon at (1/4, 0, 2/3).

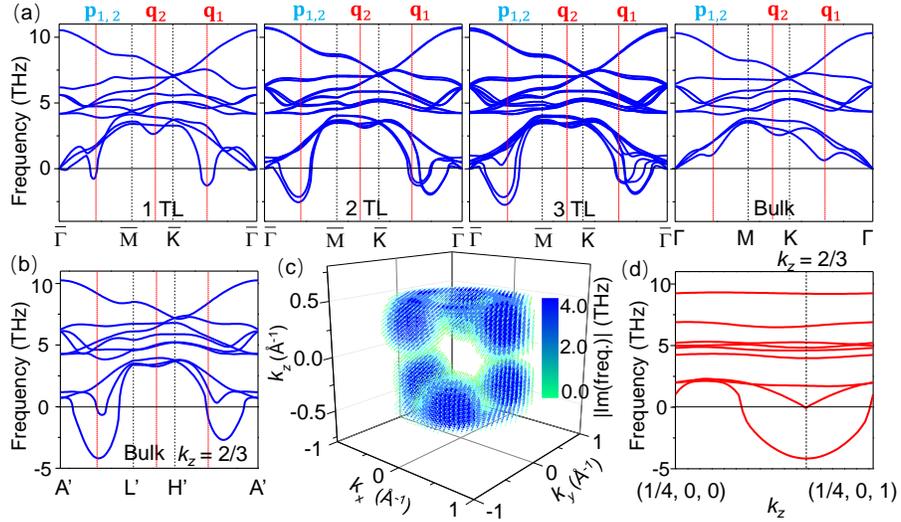